\documentclass[%
 reprint,
 superscriptaddress,
 nofootinbib,
 nobibnotes,
 amsmath,amssymb,
 aps,
 prl,
]{revtex4-2}

\usepackage{graphicx}
\usepackage{dcolumn}
\usepackage{bm}
\usepackage{braket}

\begin{document}

\title{Observation of residual entanglement in chip-based entanglement purification}

\author{Yonghe Yu}
\affiliation{Department of Electrical and Photonics Engineering, 
Technical University of Denmark, Lyngby 2800, Denmark}

\author{Mujtaba Zahidy}
\affiliation{Department of Electrical and Photonics Engineering, 
Technical University of Denmark, Lyngby 2800, Denmark}

\author{Siyan Zhou}
\affiliation{SiPhotonIC ApS, Virum 2830, Denmark}

\author{Zhongyang Wang}
\affiliation{Department of Electrical and Photonics Engineering, 
Technical University of Denmark, Lyngby 2800, Denmark}

\author{Caterina Vigliar}
\affiliation{Department of Electrical and Photonics Engineering, 
Technical University of Denmark, Lyngby 2800, Denmark}

\author{Karsten Rottwitt}
\affiliation{Department of Electrical and Photonics Engineering, 
Technical University of Denmark, Lyngby 2800, Denmark}

\author{Leif Katsuo Oxenl{\o}we}
\affiliation{Department of Electrical and Photonics Engineering, 
Technical University of Denmark, Lyngby 2800, Denmark}

\author{Yunhong Ding}
\email{yudin@dtu.dk}
\affiliation{Department of Electrical and Photonics Engineering, 
Technical University of Denmark, Lyngby 2800, Denmark}

\date{\today}

\begin{abstract}
Entanglement purification is an essential component of quantum repeaters, as it can improve the fidelity of the distributed entangled states and mitigate the effects of the noisy channel. 
Successful purification yields entangled states with increased fidelity, whereas failed events can still retain residual entanglement that remains usable for further purification when the error rates of the two degrees of freedom (DOFs) are unbalanced.
In this paper, we demonstrate a single-copy entanglement purification scheme based on hyperentanglement using silicon chips and experimentally observe the presence of residual entanglement. 
Leveraging the reconfigurability of integrated photonics, our scheme ensures that, under bit-flip noise acting on the two DOFs, residual entanglement suitable for further purification can always be obtained, regardless of which DOF has the higher error rate.
Our results demonstrate the advantages of integrated photonics for quantum information processing and provide guidance for the optimized utilization of entanglement resources in on-chip entanglement purification and future quantum repeater systems.

\end{abstract}

\maketitle


\noindent\textit{Introduction---}A primary objective of quantum communication is the distribution of entanglement \cite{bouwmeester1997experimental,briegel1998quantum,sangouard2011quantum,pan2019qure,azuma2023quantum,puriandrepea1999}, while channel noise inevitably degrades its quality during transmission \cite{werner1989quantum}. Entanglement purification can distill high-fidelity entangled states from noise-polluted ones \cite{puri41996,puri12001,puri22003,purichen2017,puriraniner2006,purikalb2017,puriKrastanov2019,yan2022entanglement} and is a central component of quantum repeaters \cite{puriandrepea1999}. In particular, hyperentanglement-based purification schemes\cite{puripan32002,puriwang2018,puri32021}, which utilize multiple degrees of freedom (DOFs) of a single photon pair, offer higher purification efficiency than conventional two-copy protocols \cite{puri41996,puri12001,puri22003}.

Typically, only states from successful purification events are preserved, while failed cases are discarded \cite{puri12001,puripan32002,puri22003}. However, recent studies have shown that when the fidelities of the two DOFs involved in entanglement purification are unequal, failed events may still contain residual entanglement \cite{zhou2020residual,zhou2025observation}. 
Such situations can arise in hyperentanglement schemes, where different DOFs, such as polarization and spatial modes, are affected by different noise processes and therefore have unequal fidelities \cite{puri32021,yu2025chip}.

Specifically, when bit-flip (BF) noise dominates and the polarization fidelity is higher than that of the spatial mode, the fidelity of the residual entanglement can still exceed 0.5 and can be further purified \cite{zhou2025observation}. Utilizing this effect provides a practical way to enhance the overall yield of entanglement purification.

Integrated silicon photonics has emerged as a promising platform for quantum information processing, owing to its advantages in high integration, scalability, stability, and compatibility with complementary metal-oxide-semiconductor (CMOS) technology \cite{pelucchi2022potential,wang2020integrated}. In addition, the inherent reconfigurability of photonic circuits allows flexible implementation of different quantum protocols without modifying the physical setup \cite{adcock2019programmable,llewellyn2020chipswapping}, making it well suited for large-scale and programmable quantum systems. On-chip entanglement purification has therefore become an important step toward scalable quantum repeaters \cite{yu2025chip}.

\begin{figure}[h]
		\centering\mbox{%
			\includegraphics[width=8cm]{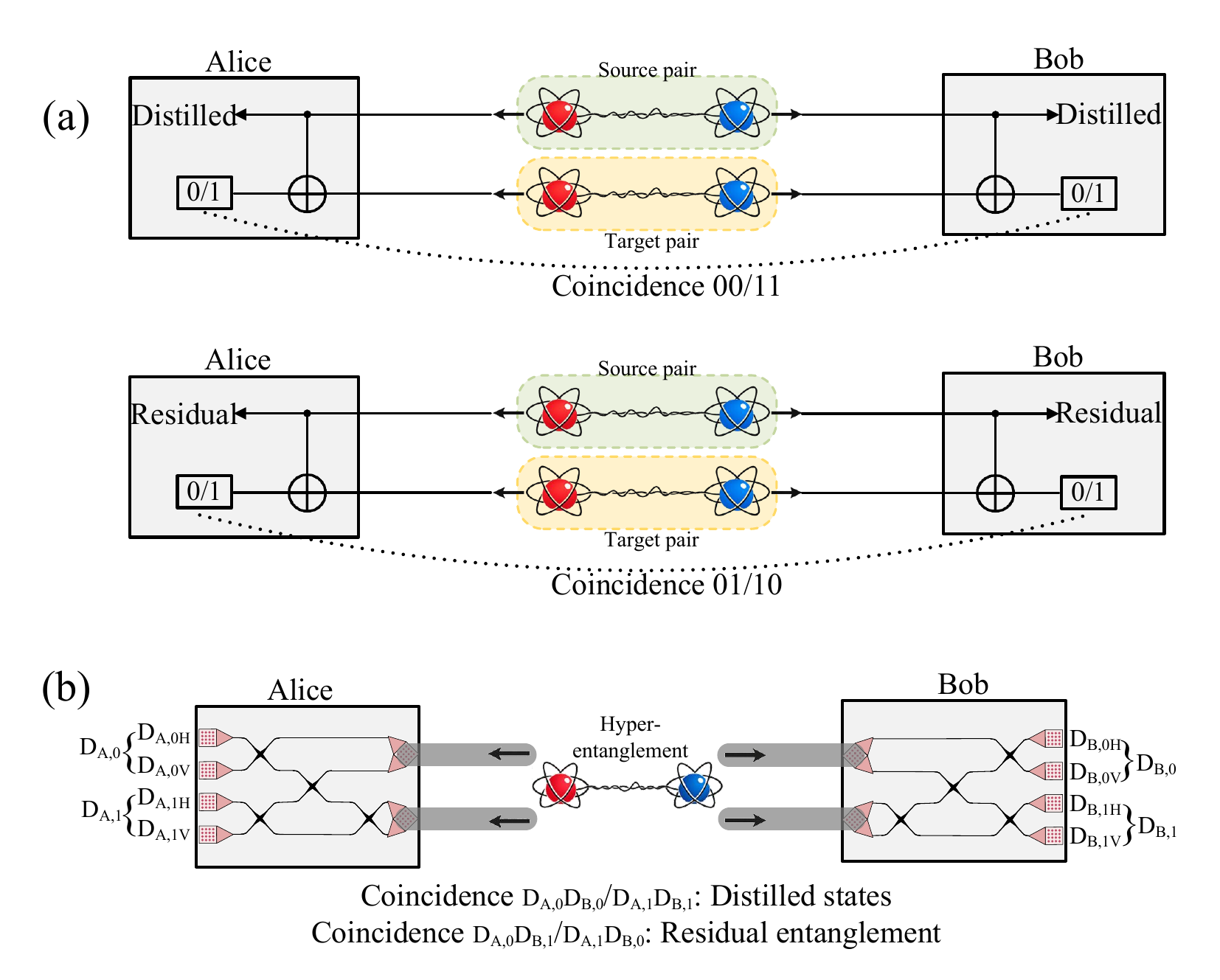}%
		}\caption{\label{fig1} Observation of residual entanglement in entanglement purification experiments. (a) Collection of distilled entanglement and residual entanglement in the purification process. (b) Chip-based entanglement purification experiment based on hyperentanglement \cite{yu2025chip}. The hyperentangled photon pairs are distributed into two pairs of optical fibers, where flip errors are introduced. Alice and Bob receive the hyperentangled states and map them onto high-dimensional entanglement on chip. The entanglement is then processed by on-chip purification circuits composed of waveguide crossings. The distilled states and residual entanglement are postselected according to the coincidence detection events.
        }
	\end{figure}

\begin{figure*}[t]
		\centering\mbox{%
			\includegraphics[width=15cm]{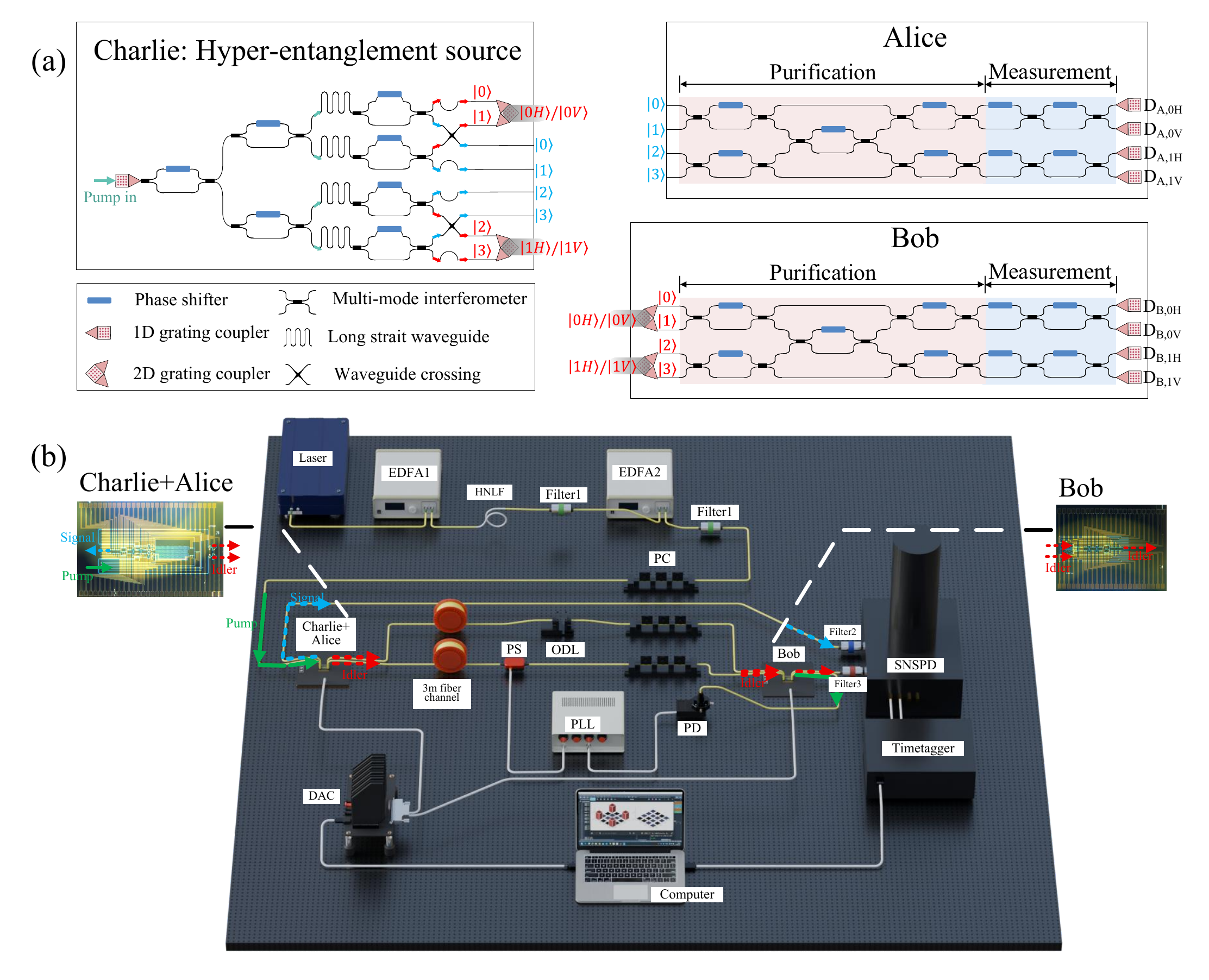}%
		}\caption{\label{fig2} The chip layout and experimental setup for chip-based entanglement purification and residual entanglement observation. 
        (a) The layout of the chips used in this work, including the hyperentanglement source and the purification circuit of Alice and Bob. Entangled photon pairs are generated by spontaneous four-wave mixing (SFWM). The signal photons keep the high-dimensional encoding and are routed to Alice through waveguides, while the idler photons are converted into polarization-spatial hyper-encoding and delivered to Bob via single-mode fibers. The Mach-Zehnder interferometers (MZI) on Alice and Bob can be configured to implement the purification protocol shown in Fig.~\ref{fig1}(b) or the identity transformation.
        (b) Experimental setup for on-chip entanglement purification. The pump is generated by a pulsed laser source, filtered at 1549.32 nm, and coupled into the chip via 1D grating couplers (GCs). The signal and idler photons after purification are coupled out of the chip also using 1D GCs. All on-chip thermo-optic phase shifters are controlled by a multi-channel digital-to-analog converter (DAC). The detection efficiency of the superconducting nanowire single-photon detectors (SNSPDs) is approximately 90\%. A phase-locked loop (PLL) system is used to stabilize the phase difference between the two 3-meter single-mode fibers carrying the idler photons.
        }
	\end{figure*}
    
In this work, we experimentally observe residual entanglement in an on-chip single-copy purification scheme. Moreover, under BF noise, by reconfiguring the silicon circuit, residual entanglement with fidelity always larger than 0.5 can be obtained even when the fidelity of spatial mode is higher, enabling more efficient use of entanglement resources in on-chip quantum repeater.


\noindent\textit{Experimental setup---}In early entanglement purification protocols \cite{puri12001,puri22003}, two identical photon pairs are typically used as input, one referred to as the source pair and the other as the target pair, as shown in Fig.~\ref{fig1}(a). The CNOT operations are applied from the source pair to the target pair, and the measurement results of the target pair indicate whether the purification succeeds or fails. If the measurement result of the two photons of the target pair are the same ($00$ or $11$), the purification succeeds and the fidelity of the source pair is improved. If the results are different ($01$ or $10$), the purification fails and the source pair is regard as not entangled. 
Recent studies have shown that when the source and target pairs are not identical \cite{zhou2020residual}, the source pair may still retain entanglement even in the failure cases. In this situation, although the target pair is measured as $01$ or $10$, the fidelity of the source pair can still be higher than $0.5$ \cite{zhou2020residual}, allowing it to be further purified in subsequent rounds.

\begin{figure*}[t]
		\centering\mbox{%
			\includegraphics[width=\textwidth]{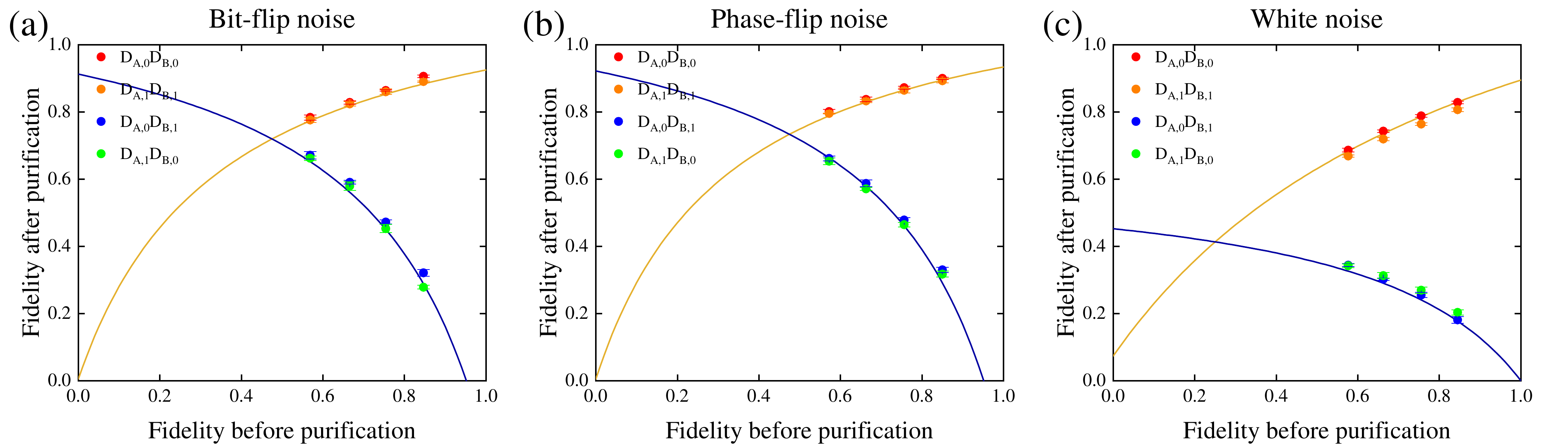}%
		}\caption{\label{fig3} Experimental results of the fidelity of the distilled entangled state and the residual entanglement under BF noise (a), PF noise (b), and white noise (c). As we explained in Fig.~\ref{fig1}(b), coincidence events $D_{A,0}D_{B,0}$ or $D_{A,1}D_{B,1}$ correspond to the distilled state, while $D_{A,0}D_{B,1}$ or $D_{A,1}D_{B,0}$ indicate purification failure, where residual entanglement may exist. In this figure, to reveal the residual entanglement, the fidelity of the spatial qubit is fixed, while different levels of noise are applied to the polarization qubit. The theoretical calculated fidelity of the distilled and residual entanglement as a function of $F_p$ is also provided.
        }
	\end{figure*}

With the assistance of hyperentanglement, residual entanglement in entanglement purification has been experimentally observed \cite{zhou2025observation}. In hyperentanglement-based scheme, the spatial-mode qubit serves as the target pair, and its measurement results indicate whether the polarization qubit is purified \cite{puripan32002}. 

To study residual entanglement in the on-chip entanglement purification experiment \cite{yu2025chip}, we group the detectors of Alice and Bob in pairs, as shown in Fig.~\ref{fig1}(b). For example, $\mathrm{D}_{A,0H}$ and $\mathrm{D}_{A,0V}$ are denoted as $\mathrm{D}_{A,0}$, corresponding to spatial mode $0$ on Alice’s side. 
In this scheme, the polarization-spatial hyperentanglement in fibers is converted into high-dimensional path-encoded entanglement on chip using 2D grating couplers (GC) and subsequently purified by the on-chip circuit.
Therefore, coincidence events $\mathrm{D}_{A,0}\mathrm{D}_{B,0}$ or $\mathrm{D}_{A,1}\mathrm{D}_{B,1}$ indicate successful purification, while $\mathrm{D}_{A,0}\mathrm{D}_{B,1}$ or $\mathrm{D}_{A,1}\mathrm{D}_{B,0}$ correspond to purification failure, where residual entanglement may still exist.

Two silicon photonic chips are used in this experiment. The first chip integrates Charlie for entanglement generation and Alice for purification, while the second chip is used for Bob’s purification. 
As shown in Fig.~\ref{fig2}(a), in Charlie’s circuit, the pump light is coupled in through a 1D GC and distributed into four long straight waveguides, where spontaneous four-wave mixing (SFWM) occurs. In each pump pulse, one signal-idler photon pair is generated probabilistically across the four waveguides. After demultiplexing with asymmetric Mach-Zehnder interferometers (MZIs), the signal and idler photons are both encoded in four path modes, labeled as $\ket{0}$, $\ket{1}$, $\ket{2}$, and $\ket{3}$, corresponding to the waveguide in which the photon is present. The resulting entangled state can be written as $\frac{1}{2}(\ket{00} + \ket{11} + \ket{22} + \ket{33})$.
The signal photons are directly routed to Alice through waveguides for purification (red background box) and subsequent measurement (blue background box). The idler photons are converted from high-dimensional encoding on chip into polarization-spatial hyper-encoding \cite{yu2025chip} in fibers using 2D GCs and are then sent to Bob through two single-mode fibers. Bob performs the same purification operations as Alice, followed by measurement. The measurement circuits outputs from both Alice and Bob are coupled off chip through 1D GCs and detected by off-chip superconducting nanowire single-photon detectors (SNSPDs).

As shown in Fig.~\ref{fig2}(b), the pump light is generated by a mode-locked pulsed laser with a repetition rate of 10 GHz and amplified by erbium-doped fiber amplifiers (EDFA), and then it is filtered with a 200 GHz filter at 1549.32 nm (Filter1). A phase shifter (PS) and an optical delay line (ODL) are inserted in the two optical fibers connecting Charlie and Bob to stabilize the phase and compensate the path-length difference.
On Bob’s chip, the circuit converts the phase difference between the two optical fibers into an intensity variation \cite{yu2025chip}, which is detected by an off-chip photodiode (PD). The signal from PD is processed by a phase-locked loop (PLL), which provides feedback to the PS to stabilize the phase. To suppress the pump light noise from the signal and idler photons, 100 GHz bandpass filters centered at 1539.77 nm for signal (Filter2) and 1558.98 nm for idler (Filter3) are placed before the SNSPDs. All on-chip thermo-optic phase shifters are controlled by a multi-channel digital-to-analog converter (DAC), which sets the heater voltages to tune the phase.

In this work, on-chip circuits are used to simulate noise in optical fibers \cite{yu2025chip}, including bit-flip (BF) and phase-flip (PF) errors (see Supplemental Material). The Hadamard gate for the purification of PF errors is also implemented with on-chip circuits \cite{yu2025chip}. By reconfiguring the purification circuits on Alice and Bob and postselecting coincidence events according to the method described in Fig.~\ref{fig1}(b), both entanglement purification and residual entanglement can be observed.


\noindent\textit{Experimental result---}In Fig.~\ref{fig3}(a), BF errors are applied to both the polarization and spatial qubits. The fidelity of the spatial qubit is fixed at $F_s = 0.749 \pm 0.009$, while the fidelity of the polarization qubit before purification $F_p$ is varied from $0.569 \pm 0.003$ to $0.847 \pm 0.003$. 
Coincidence events $D_{A,0}D_{B,0}$ (red circles) and $D_{A,1}D_{B,1}$ (orange circles) correspond to the distilled states after purification, while $D_{A,0}D_{B,1}$ (blue circles) and $D_{A,1}D_{B,0}$ (green circles) correspond to residual entanglement. For the distilled states, as shown in Fig.~\ref{fig3}(a), the fidelity of is improved for all $F_p$. 
For the residual entanglement, when $F_p < F_s$, the fidelity remains above $0.5$ and can be further purified in the next round of purification \cite{zhou2020residual}. When $F_p > F_s$, for example at $F_p = 0.847 \pm 0.003$, the fidelity of the residual entanglement is $0.321 \pm 0.011$, which is below $0.5$ \cite{zhou2020residual}. At the transition point $F_p = 0.754 \pm 0.005$, the residual fidelity is $0.472 \pm 0.006$, slightly below the theoretical value of $0.5$, which may be attributed to imperfect spectral purity and background noise.

For the case with only PF errors, Hadamard operations are applied on both the polarization and spatial qubits to convert them into BF errors. The states are then purified by our circuit, with results similar to those under BF noise \cite{zhou2025observation}. The result is shown in Fig.~\ref{fig3}(b), where the spatial fidelity is fixed at $F_s = 0.755 \pm 0.006$. As the polarization fidelity before purification increases from $0.572 \pm 0.004$ to $0.851 \pm 0.007$, the fidelity of the residual entanglement decreases from $0.662 \pm 0.006$ to $0.330 \pm 0.008$ after purification.

For white noise, both the spatial and polarization qubits are prepared in Werner states \cite{werner1989quantum}. In Fig.~\ref{fig3}(c), the spatial fidelity is set to $F_s = 0.734 \pm 0.008$, and the initial polarization fidelity is varied from $0.576 \pm 0.006$ to $0.845 \pm 0.004$. 
The fidelity of the distilled state varies from $0.686 \pm 0.006$ to $0.828 \pm 0.004$, indicating that the fidelity can decrease even in successful purification events. The fidelity of the residual entanglement decreases from $0.344 \pm 0.004$ to $0.181 \pm 0.010$, indicating that the residual entanglement is not usable for further purification under white noise \cite{zhou2020residual}.

\begin{figure}[h]
\centering
\includegraphics[width=9cm]{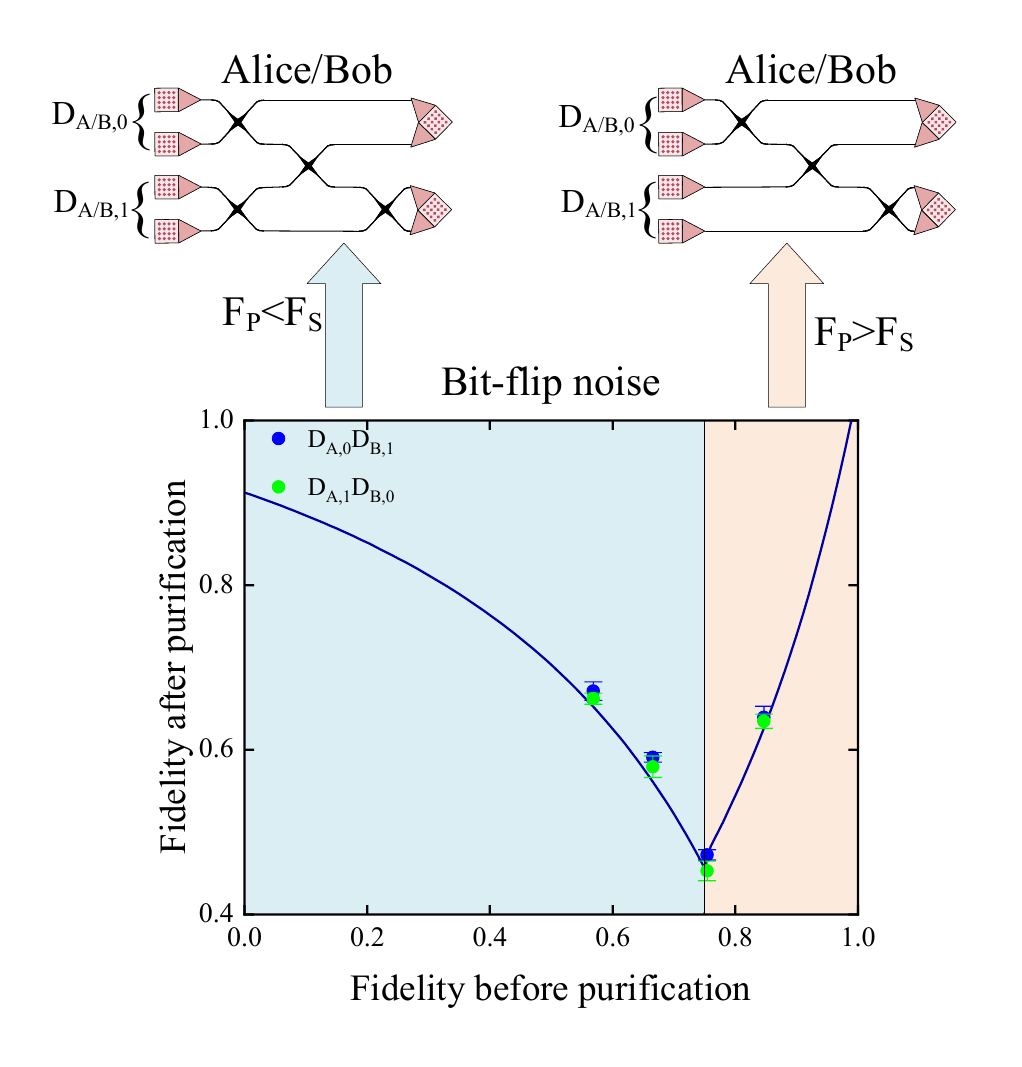}
\caption{Experimental demonstration of the residual entanglement as a function of the polarization-state fidelity $F_p$ under different circuit configurations. For $F_p < F_s$, the original purification circuit is applied (light blue arrow), whereas for $F_p > F_s$, the reconfigured circuit indicated by the light red arrows is used. The experimental results for the residual entanglement after purification also show good agreement with the theoretical calculation.}\label{fig4}
\end{figure}


An important point is how to fully utilize the residual entanglement, particularly when the fidelity falls below $0.5$ in Fig.~\ref{fig3}. In Fig.~\ref{fig4}, we demonstrate a strategy under BF noise that ensures the fidelity of the residual entanglement remains above $0.5$ over the entire range of $F_p$.
For the case $F_p < F_s$, we apply the original purification circuit in Alice and Bob, and the residual fidelity decreases to 0.5 as $F_p$ increase approaches $F_s$.
For $F_p > F_s$, the MZIs before $D_{A,1}$ and $D_{B,1}$ are reconfigured to implement the identity operation rather than the waveguide crossing, as shown in Fig.~\ref{fig4}. With this circuit (indicated by the light red arrow), the decreasing trend of the residual fidelity is reversed, reaching $0.640 \pm 0.013$ at $F_p = 0.847 \pm 0.003$.
This behavior arises because, under BF noise, the residual state is a mixture of $\ket{\Phi^+} = \frac{1}{\sqrt{2}} \left( \ket{00} + \ket{11} \right)$ with weight $F_s(1-F_p)/N$ and $\ket{\Psi^+} = \frac{1}{\sqrt{2}} \left( \ket{01} + \ket{10} \right)$ with weight $F_p(1-F_s)/N$. For $F_p > F_s$, the $\ket{\Psi^+}$ component dominates. By reconfiguring the circuit to flip this component back to $\ket{\Phi^+}$, a higher fidelity is obtained. Free-space purification schemes require additional optical components (e.g., HWPs) to implement this transformation~\cite{zhou2025observation}, whereas here it is readily achieved through circuit reconfiguration.

This operation does not affect the distilled states. It does not modify the outcomes corresponding to $D_{A,0}D_{B,0}$, and for $D_{A,1}D_{B,1}$ it is equivalent to applying BF operations on the signal and idler photons separately, leaving the state unchanged.

\noindent\textit{Conclusion---}In this work, we experimentally observe residual entanglement in on-chip entanglement purification \cite{yu2025chip}, providing guidance for the efficient utilization of entanglement resources in chip-based quantum networks. The purification scheme in this work is designed to convert hyperentanglement in optical fibers into high-dimensional path-encoded entanglement on chip and to mitigate BF and PF noise arising in fiber transmission using on-chip quantum circuits. We investigate three noise models: BF, PF, and white noise. For BF and PF noise, the fidelity of the distilled states is improved, while residual entanglement can be observed when $F_p < F_s$. In contrast, under white noise, no residual entanglement is observed regardless of the fidelity of the state before purification.

We further experimentally demonstrate a strategy to fully utilize residual entanglement under BF noise by reconfiguring on-chip circuit, even when $F_p > F_s$. Such behavior cannot be observed in free-space schemes without introducing additional components~\cite{zhou2025observation}. This implies that when the fidelities of the source and target qubits differ, residual entanglement always exists \cite{zhou2020residual}. Such imbalance is inherent in single-copy purification schemes based on different DOFs, as each DOF experiences distinct noise processes \cite{puri32021}. Our approach therefore provides a practical route for on-chip hyperentanglement purification and enhances the efficient utilization of entanglement resources.

\vspace{0.5em}
\begin{acknowledgments}
\noindent\textit{    Acknowledgments---}We acknowledge funding from Villum Fonden Young Investigator project QUANPIC (Ref. 00025298) and Danish National Research Foundation Center of Excellence, SPOC (Ref. DNRF123). The authors thank Yu-Bo Sheng for helpful discussions.
\end{acknowledgments}


\bibliography{apssamp}

\clearpage 
\onecolumngrid
\newcommand{\papertitle}[2]{%
  \begin{center}
    {\LARGE #1\par} 
  \end{center}
  \vspace{1em}
}
\papertitle{Supplementary Information: Observation of residual entanglement in chip-based entanglement purification}

\setcounter{page}{1}
\setcounter{section}{0}
\setcounter{figure}{0}
\setcounter{table}{0}
\setcounter{equation}{0}

\renewcommand{\thefigure}{S\arabic{figure}}
\renewcommand{\thetable}{S\arabic{table}}
\renewcommand{\theequation}{S\arabic{equation}}

\makeatletter
\renewcommand{\theHfigure}{paperB.S\arabic{figure}}
\renewcommand{\theHtable}{paperB.S\arabic{table}}
\renewcommand{\theHequation}{paperB.S\arabic{equation}}
\renewcommand{\theHsection}{paperB.sec.\arabic{section}}
\makeatother

\section{Chip and device details}

The chips are fabricated on 250 nm silicon-on-insulator (SOI) platform using deep ultraviolet (DUV) lithography in a muti-project (MPW) wafer run by SiPhotonIC ApS, and subsequently packaged on a printed circuit board. Two fiber arrays are aligned with the on-chip grating couplers and fixed on each chip, with all 2D GCs located on the left side and all 1D GCs on the right side. The coupling angles of the 1D and 2D GCs are optimized to achieve coupling efficiencies of $-5.3$ dB at 1550 nm for the 1D GCs and $-5.9$ dB at 1560 nm for the 2D GCs~\cite{xue2019two} after packaging.

In the experiment, with a pump power of 18.9 dBm before the chip, a coincidence rate of 10.38 Hz and a coincidence-to-accidental ratio (CAR) of 56.3 are obtained. The corresponding squeezing parameter is $\xi = 0.02$, indicating that the multi-photon-pair contribution is negligible. The spectral purity is then estimated to be 0.94.

\section{Noise simulation}
\begin{figure}[h]
    \centering
    \includegraphics[width=\textwidth]{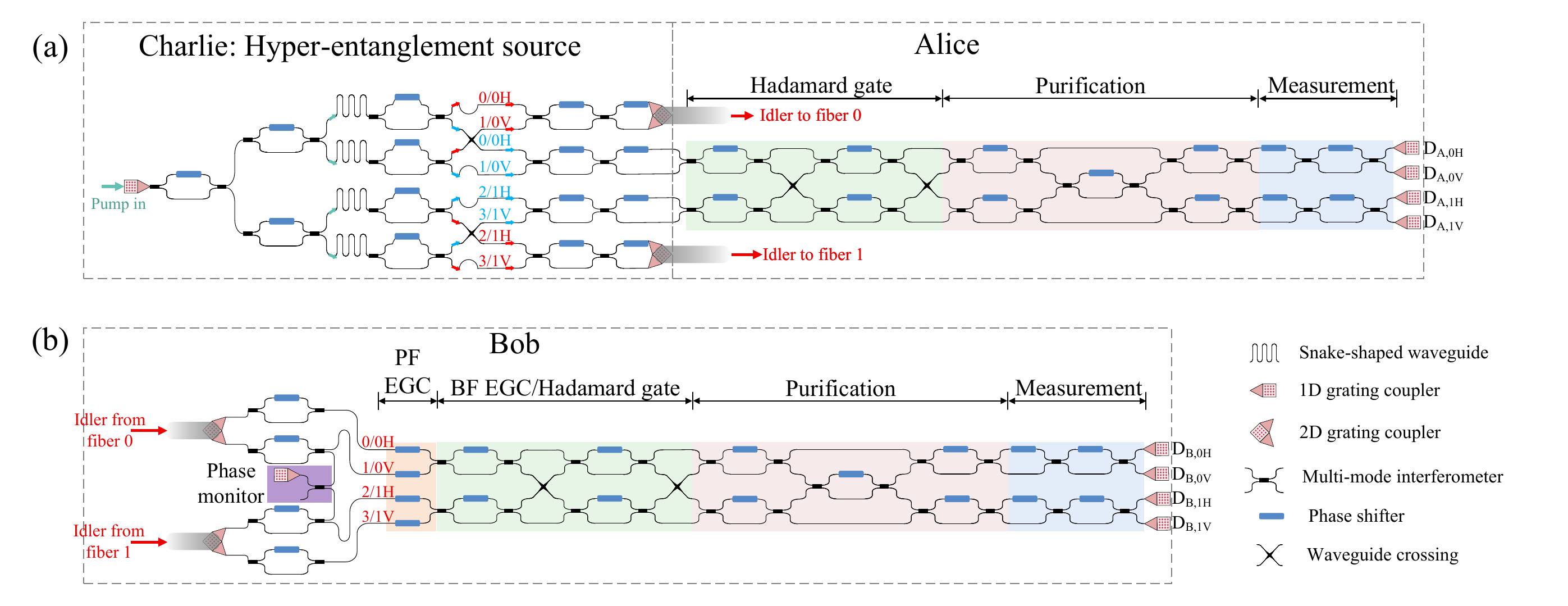}
    \caption{Full circuit layouts of Chip Charlie/Alice (a) and Chip Bob (b).}
    \label{supp1}
\end{figure}
As mentioned above, two silicon photonic chips are used in this experiment. The full layout is shown in Fig.~\ref{supp1}. 
To compensate for the power imbalance among the four waveguides converted from the two fibers, four MZIs are placed after the 2D grating couplers \cite{llewellyn2020chipswapping}. These MZIs also split the optical power and interfere the signals in the phase-monitoring circuit, which convert the phase difference between the two fibers into an intensity variation.
After that, the noise-simulation circuits for BF and PF noise are implemented on Bob’s chip, before the purification circuit.
By introducing extra phases on the four phase shifters in the PF EGC (orange box), PF errors are implemented on the path-encoded qubit, equivalent to the PF errors on polarization and spatial qubits arising in optical fibers~\cite{yu2025chip}. 
By reconfiguring the four MZIs in the BF EGC (green box), BF errors generated in fiber transmission can be simulated~\cite{yu2025chip}.

In addition, to purify PF errors, Hadamard operations are applied on both Alice’s and Bob’s sides to convert them into BF errors \cite{yu2025chip}. Accordingly, the Hadamard gate circuit (green box) is implemented on Alice’s chip, while the MZIs in the BF EGC on Bob’s chip can also be reconfigured to realize the Hadamard gate operation.

\begin{figure}[h]
    \centering
    \includegraphics[width=\textwidth]{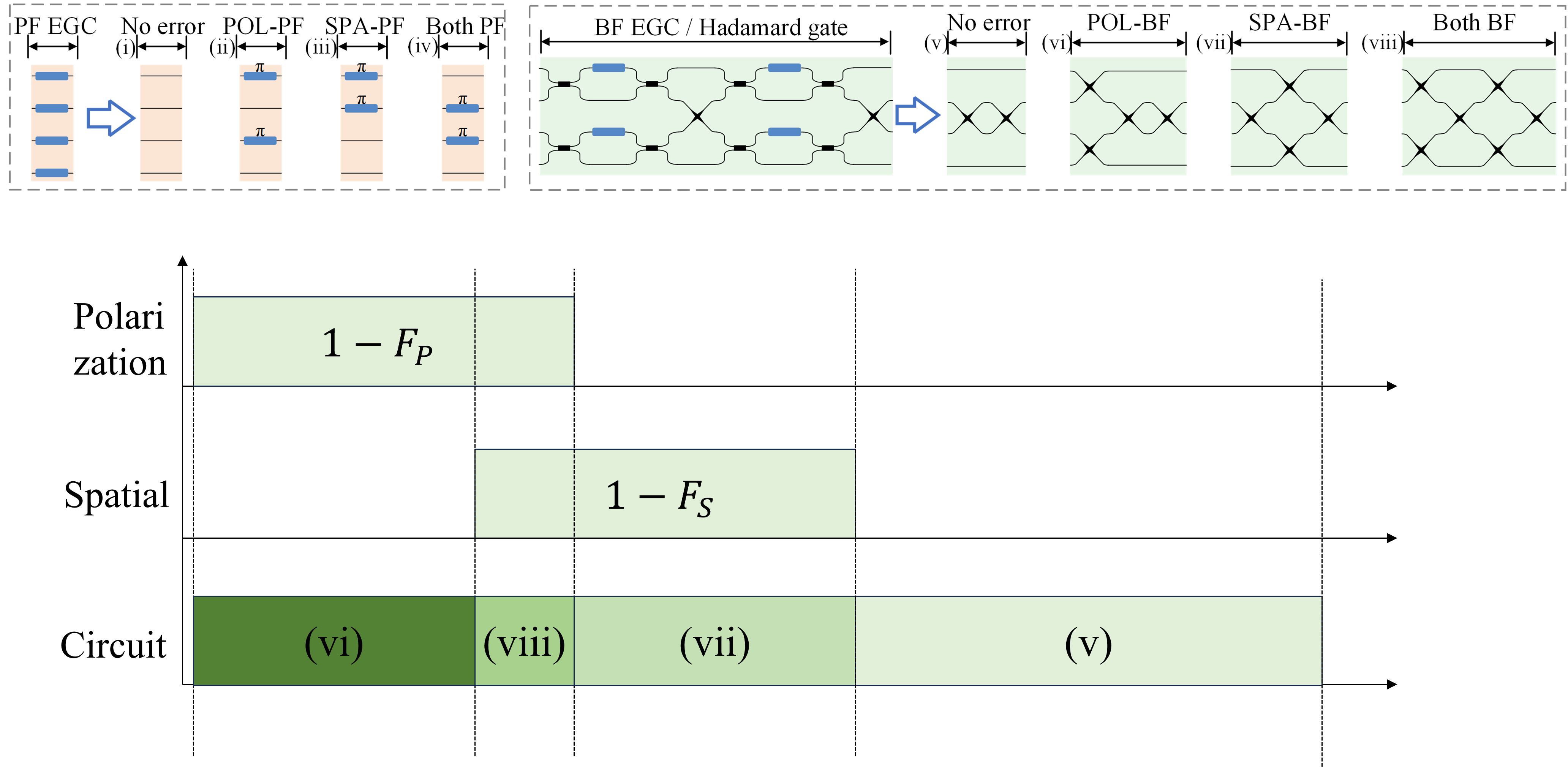}
    \caption{The method we simulate the BF error in a time distribution \cite{yu2025chip}. The fidelitys of the spatial and polarization qubit are $F_s$ and $F_p$, respectively.}
    \label{timedistrerror}
\end{figure}

The BF EGC at Bob's chip is reconfigured to the circuits as shown in Fig.~\ref{timedistrerror}(v)-\ref{timedistrerror}(viii) in a time distribution. In general, one time window is divided into four time slots, each corresponding to a specific circuit configuration. The duration of each time slot is determined according to the BF error rates on the polarization and spatial qubits~\cite{puri32021}.

\begin{figure}[h]
    \centering
    \includegraphics[width=13cm]{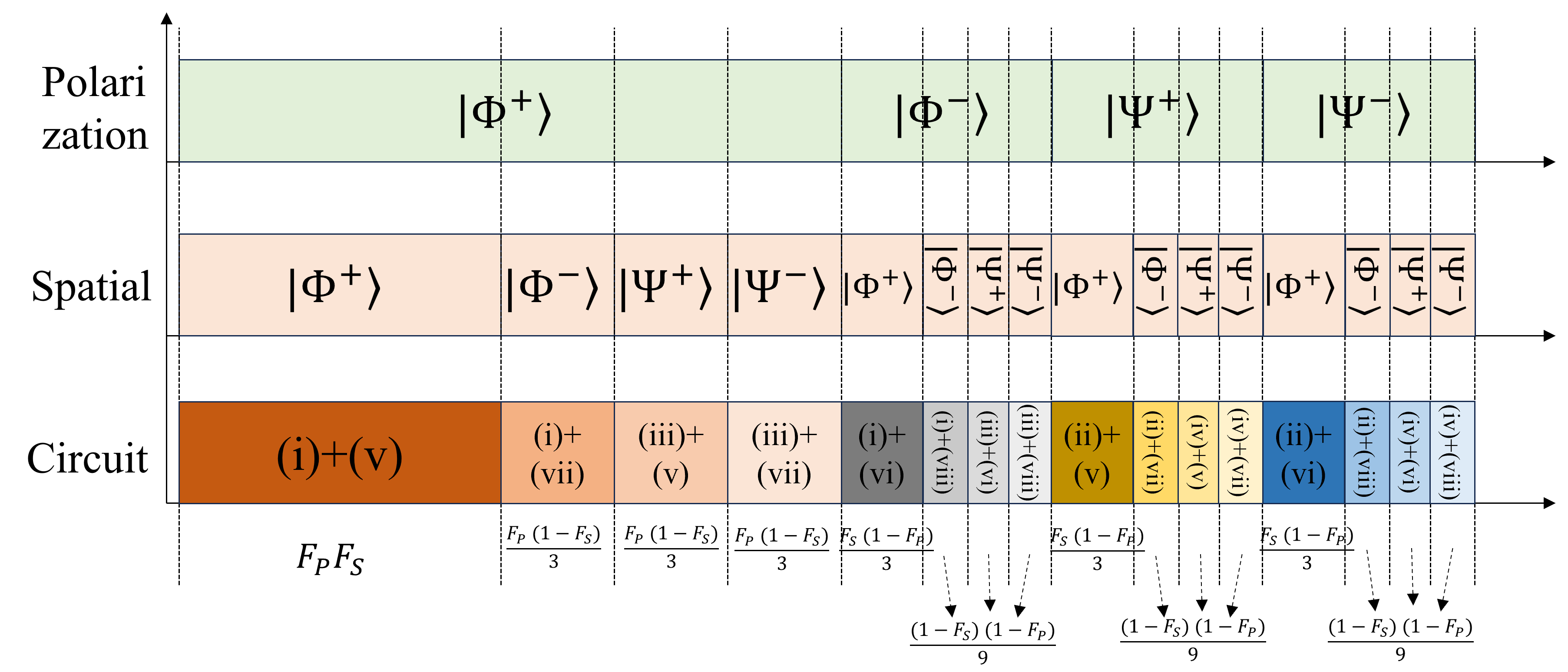}
    \caption{The method we simulate the white noise in a time distribution. }
    \label{timedistrerror2}
\end{figure}

The PF error simulation follows a similar procedure as described in Fig.~\ref{timedistrerror}. For white noise, the full time window is divided into 16 time slots, corresponding to all possible combinations of two-qubit basis states. The corresponding time distribution is shown in Fig.~\ref{timedistrerror2}.

\section{Theoretical model of entanglement purification and residual entanglement}

In this part, we assume that the state after transmission through the fibers is a hyperentangled state $\rho = \rho_s \otimes \rho_p$, where
\begin{equation}
\begin{aligned}
\rho_s &= F_1 \ket{\phi^+}\bra{\phi^+} + A_1 \ket{\phi^-}\bra{\phi^-} + B_1 \ket{\psi^+}\bra{\psi^+} + C_1 \ket{\psi^-}\bra{\psi^-}, \\
\rho_p &= F_2 \ket{\Phi^+}\bra{\Phi^+} + A_2 \ket{\Phi^-}\bra{\Phi^-} + B_2 \ket{\Psi^+}\bra{\Psi^+} + C_2 \ket{\Psi^-}\bra{\Psi^-}.
\end{aligned}
\label{eq1}
\end{equation}

The analysis of this part has been presented in previous work~\cite{zhou2020residual}. Here, we summarize it in Table~\ref{table1} using on-chip path-encoded qubits for clarity.

\begin{table}[h]
\centering
\caption{\label{table1}The table lists the probabilities of BF and PF errors occurring and whether the errored states still cause coincidences after purification. Spatial (Before) and Polar (Before) denote the spatial-mode bit and polarization bit in the fiber and their probability, respectively. HD state (After) represents the path-encoded high-dimensional state on the chip after the 2D GCs and purification circuit.}
\resizebox{14cm}{!}{%
\begin{tabular}{@{}llllll@{}}
\hline
Spa (Before) & Pol (Before) & HD state (After) & Proba & Distilled & Residual \\
\hline
$\ket{\phi^+}\ F_1$ & $\ket{\Phi^+}\ F_2$ & $\ket{11}+\ket{33}+\ket{22}+\ket{00}$ & $F_1F_2$ & $\ket{00}+\ket{11}$ & \\
$\ket{\phi^+}\ F_1$ & $\ket{\Phi^-}\ A_2$ & $\ket{11}-\ket{33}+\ket{22}-\ket{00}$ & $F_1A_2$ & $\ket{00}-\ket{11}$ & \\
$\ket{\phi^+}\ F_1$ & $\ket{\Psi^+}\ B_2$ & $\ket{13}+\ket{31}+\ket{20}+\ket{02}$ & $F_1B_2$ & & $\ket{00}+\ket{11}$ \\
$\ket{\phi^+}\ F_1$ & $\ket{\Psi^-}\ C_2$ & $\ket{13}-\ket{31}+\ket{20}-\ket{02}$ & $F_1C_2$ & & $\ket{00}-\ket{11}$ \\[5pt]

$\ket{\phi^-}\ A_1$ & $\ket{\Phi^+}\ F_2$ & $\ket{11}+\ket{33}-\ket{22}-\ket{00}$ & $A_1F_2$ & $\ket{00}-\ket{11}$ & \\
$\ket{\phi^-}\ A_1$ & $\ket{\Phi^-}\ A_2$ & $\ket{11}-\ket{33}-\ket{22}+\ket{00}$ & $A_1A_2$ & $\ket{00}+\ket{11}$ & \\
$\ket{\phi^-}\ A_1$ & $\ket{\Psi^+}\ B_2$ & $\ket{13}+\ket{31}-\ket{20}-\ket{02}$ & $A_1B_2$ & & $\ket{00}-\ket{11}$ \\
$\ket{\phi^-}\ A_1$ & $\ket{\Psi^-}\ C_2$ & $\ket{13}-\ket{31}-\ket{20}+\ket{02}$ & $A_1C_2$ & & $\ket{00}+\ket{11}$ \\[5pt]

$\ket{\psi^+}\ B_1$ & $\ket{\Phi^+}\ F_2$ & $\ket{12}+\ket{30}+\ket{21}+\ket{03}$ & $B_1F_2$ & & $\ket{01}+\ket{10}$ \\
$\ket{\psi^+}\ B_1$ & $\ket{\Phi^-}\ A_2$ & $\ket{12}-\ket{30}+\ket{21}-\ket{03}$ & $B_1A_2$ & & $\ket{01}-\ket{10}$ \\
$\ket{\psi^+}\ B_1$ & $\ket{\Psi^+}\ B_2$ & $\ket{10}+\ket{32}+\ket{23}+\ket{01}$ & $B_1B_2$ & $\ket{01}+\ket{10}$ & \\
$\ket{\psi^+}\ B_1$ & $\ket{\Psi^-}\ C_2$ & $\ket{10}-\ket{32}+\ket{23}-\ket{01}$ & $B_1C_2$ & $\ket{01}-\ket{10}$ & \\[5pt]

$\ket{\psi^-}\ C_1$ & $\ket{\Phi^+}\ F_2$ & $\ket{12}+\ket{30}-\ket{21}-\ket{03}$ & $C_1F_2$ & & $\ket{01}-\ket{10}$ \\
$\ket{\psi^-}\ C_1$ & $\ket{\Phi^-}\ A_2$ & $\ket{12}-\ket{30}-\ket{21}+\ket{03}$ & $C_1A_2$ & & $\ket{01}+\ket{10}$ \\
$\ket{\psi^-}\ C_1$ & $\ket{\Psi^+}\ B_2$ & $\ket{10}+\ket{32}-\ket{23}-\ket{01}$ & $C_1B_2$ & $\ket{01}-\ket{10}$ & \\
$\ket{\psi^-}\ C_1$ & $\ket{\Psi^-}\ C_2$ & $\ket{10}-\ket{32}-\ket{23}+\ket{01}$ & $C_1C_2$ & $\ket{01}+\ket{10}$ & \\
\hline
\end{tabular}%
}
\end{table}

From Table~\ref{table1}, the fidelities of the distilled state $F_3$ and the residual entanglement $F_4$ are obtained by summing the probabilities of the $\ket{00}+\ket{11}$ components. The resulting expressions are given by
\begin{equation}
\begin{aligned}
F_3 &= \frac{F_1F_2 + A_1A_2}{(F_1 + A_1)(F_2 + A_2) + (B_1 + C_1)(B_2 + C_2)}, \\
F_4 &= \frac{F_1B_2 + A_1C_2}{(F_1 + A_1)(B_2 + C_2) + (F_2 + A_2)(B_1 + C_1)}.
\end{aligned}
\label{eq2}
\end{equation}

\begin{figure}[h]
    \centering
    \includegraphics[width=8cm]{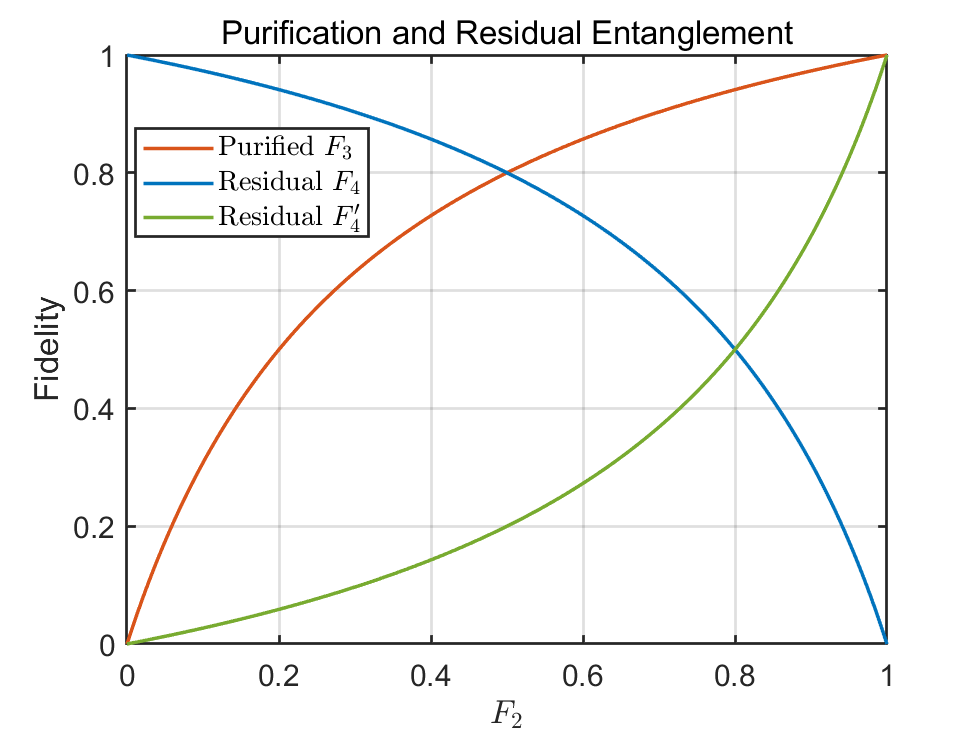}
    \caption{Fidelities of the distilled state ($F_3$, red) and the residual state with respect to $\ket{00}+\ket{11}$ ($F_4$, blue) and $\ket{01}+\ket{10}$ ($F_4'$, green). $F_2$ is the fidelity of the polarization qubit before purification.}
    \label{supp4}
\end{figure}

When there is only BF errors, we have $A_1 = C_1 = A_2 = C_2 = 0$, $B_1 = 1 - F_1$, and $B_2 = 1 - F_2$. In this case, the fidelity of the residual entanglement with respect to $\ket{00}+\ket{11}$ is $F_4 = \frac{F_1(1-F_2)}{F_1(1-F_2) + F_2(1-F_1)}$. The fraction of the $\ket{01}+\ket{10}$ component is $F_4' = \frac{F_2(1-F_1)}{F_1(1-F_2) + F_2(1-F_1)}$. Here we show a theoretical simulation of the fractions of $\ket{00}+\ket{11}$ and $\ket{01}+\ket{10}$ components when $F_1 = 0.2$ in Fig.~\ref{supp4}.

From Fig.~\ref{supp4}, we see that when the spatial fidelity $F_1$ is higher than the polarization fidelity $F_2$ ($F_2 < 0.8$), the $\ket{00}+\ket{11}$ component dominates the residual entanglement. When $F_2 > 0.8$, the $\ket{01}+\ket{10}$ component becomes dominant. 
Specially, in the extreme case $F_2 = 1$, where the polarization qubit has no BF errors, the residual entanglement consists entirely of the $\ket{01}+\ket{10}$ component.
By applying the circuit indicated by the light red arrow in Fig.~\ref{fig4}, the $\ket{01}+\ket{10}$ component is effectively used as the target state for evaluating the fidelity, leading to an increasing trend even for $F_2 > 0.8$.

\section{Parameter estimation}

In the experiment, background noise, imperfect spectral purity, inaccurate phase control of the MZIs, and insufficient filtering of the pump photons all contribute to unexpected coincidence counts.
Thus when fitting the data in Fig.~\ref{fig3}, even under BF and PF noise, all four Bell state components should be included in the mixed state, similar to Eq.~\ref{eq1}. For example, in the case of BF noise, the state is modeled as:

\begin{equation}
\begin{aligned}
\rho_s &= F_s \ket{\phi^+}\bra{\phi^+} + A_s \ket{\phi^-}\bra{\phi^-} + B_s \ket{\psi^+}\bra{\psi^+} + C_s \ket{\psi^-}\bra{\psi^-}, \\
\rho_p &= F_p \ket{\Phi^+}\bra{\Phi^+} + A_p \ket{\Phi^-}\bra{\Phi^-} + (D_p-F_p) \ket{\Psi^+}\bra{\Psi^+} + C_p \ket{\Psi^-}\bra{\Psi^-}.
\end{aligned}
\label{eq3}
\end{equation}

In the fitting of the BF data, $F_p$ is treated as the independent variable. The parameters $F_s$, $A_s$, $B_s$, $C_s$, $A_p$, $D_p$, and $C_p$ are then estimated from the experimental data ($F_3$ and $F_4$).
The fitted parameters for the BF, PF, and white-noise models are summarized in Table~\ref{table2}.

\begin{table}[h]
\centering
\caption{Estimated parameters extracted from the experimental data.}
\label{table2}
\begin{tabular}{lcccc}
\hline
Noise type & $F$ & $A$ & $B$ & $C$ \\
\hline
BF (spatial) & 0.7446 & 0.0051 & 0.1809 & 0.0010 \\
BF (polarization) & $F_p$ & 0.0497 & $0.9690 - F_p$ & 0.0408 \\
\hline
PF (spatial) & 0.7330 & 0.0238 & 0.1499 & 0.0009 \\ 
PF (polarization) & $F_p$ & 0.0281 & $0.9729 - F_p$ & 0.0689 \\
\hline
White (spatial) & 0.7888 & 0.0932 & 0.0857 & 0.0858 \\ 
White (polarization) & $F_p$ & $\frac{1 - F_p}{3.0248}$ & $\frac{1 - F_p}{2.9619}$ & $\frac{1 - F_p}{3.0492}$ \\
\hline
\end{tabular}
\end{table}

In Table~\ref{table2}, the results are obtained after applying Hadamard operations, which leads to a dominant $B$ component corresponding to $\ket{\Psi^+}\bra{\Psi^+}$. 

For white noise, the state is modeled as
\begin{equation}
\begin{aligned}
\rho_s &= F_s \ket{\phi^+}\bra{\phi^+} + A_s \ket{\phi^-}\bra{\phi^-} + B_s \ket{\psi^+}\bra{\psi^+} + C_s \ket{\psi^-}\bra{\psi^-}, \\
\rho_p &= F_p \ket{\Phi^+}\bra{\Phi^+} + \frac{1-F_p}{D_p} \ket{\Phi^-}\bra{\Phi^-} + \frac{1-F_p}{E_p} \ket{\Psi^+}\bra{\Psi^+} + \frac{1-F_p}{G_p} \ket{\Psi^-}\bra{\Psi^-}.
\end{aligned}
\label{eq4}
\end{equation}
where $F_s$, $A_s$, $B_s$, $C_s$, $D_p$, $E_p$, and $G_p$ are fitting parameters extracted from the experimental data.

\section{Detail of experimental result}

Tables~\ref{table3}--\ref{table5} show the data before and after purification for BF, PF, and white noise.

\begin{table}[h]
\centering
\caption{Experimental data of BF noise.}
\label{table3}
\begin{tabular}{lcccc}
\hline
Spatial (before) &  \multicolumn{4}{c}{$0.7492 \pm 0.0086$}  \\

Polarization (before) 
& $0.5687 \pm 0.0032$ & $0.6657 \pm 0.0054$ 
& $0.7540 \pm 0.0051$ & $0.8466 \pm 0.0033$ \\
\hline

$D_{A,0}D_{B,0}$ 
& $0.7838 \pm 0.0079$ & $0.8282 \pm 0.0053$ 
& $0.8641 \pm 0.0023$ & $0.9061 \pm 0.0041$ \\

$D_{A,0}D_{B,1}$ 
& $0.6713 \pm 0.0111$ & $0.5908 \pm 0.0058$ 
& $0.4724 \pm 0.0063$ & $0.3211 \pm 0.0092$ \\

$D_{A,1}D_{B,1}$ 
& $0.7761 \pm 0.0076$ & $0.8240 \pm 0.0064$ 
& $0.8602 \pm 0.0080$ & $0.8903 \pm 0.0026$ \\

$D_{A,1}D_{B,0}$ 
& $0.6621 \pm 0.0068$ & $0.5794 \pm 0.0131$ 
& $0.4530 \pm 0.0118$ & $0.2786 \pm 0.0052$ \\
\hline
\end{tabular}
\end{table}

\begin{table}[h]
\centering
\caption{Experimental data of PF noise.}
\label{table4}
\begin{tabular}{lcccc}
\hline
Spatial (before) &  \multicolumn{4}{c}{$0.7549 \pm 0.0033$}  \\

Polarization (before) 
& $0.5724 \pm 0.0051$ & $0.6629 \pm 0.0047$ 
& $0.7563 \pm 0.0084$ & $0.8505 \pm 0.0074$ \\
\hline
$D_{A,0}D_{B,0}$ 
& $0.8011 \pm 0.0061$ & $0.8371 \pm 0.0068$ 
& $0.8722 \pm 0.0046$ & $0.8998 \pm 0.0033$ \\

$D_{A,0}D_{B,1}$ 
& $0.6616 \pm 0.0066$ & $0.5870 \pm 0.0100$ 
& $0.4784 \pm 0.0068$ & $0.3298 \pm 0.0077$ \\

$D_{A,1}D_{B,1}$ 
& $0.7956 \pm 0.0013$ & $0.8328 \pm 0.0052$ 
& $0.8642 \pm 0.0077$ & $0.8928 \pm 0.0066$ \\

$D_{A,1}D_{B,0}$ 
& $0.6533 \pm 0.0100$ & $0.5710 \pm 0.0050$ 
& $0.4642 \pm 0.0063$ & $0.3175 \pm 0.0092$ \\
\hline
\end{tabular}
\end{table}

\begin{table}[h]
\centering
\caption{Experimental data of white noise.}
\label{table5}
\begin{tabular}{lcccc}
\hline
Spatial (before) &  \multicolumn{4}{c}{$0.7344 \pm 0.0075$}  \\

Polarization (before) 
& $0.5763 \pm 0.0063$ & $0.6624 \pm 0.0035$ 
& $0.7553 \pm 0.0054$ & $0.8447 \pm 0.0041$ \\
\hline
$D_{A,0}D_{B,0}$ 
& $0.6862 \pm 0.0059$ & $0.7426 \pm 0.0041$ 
& $0.7883 \pm 0.0039$ & $0.8283 \pm 0.0038$ \\

$D_{A,0}D_{B,1}$ 
& $0.3441 \pm 0.0041$ & $0.3024 \pm 0.0034$ 
& $0.2552 \pm 0.0074$ & $0.1813 \pm 0.0098$ \\

$D_{A,1}D_{B,1}$ 
& $0.6690 \pm 0.0038$ & $0.7193 \pm 0.0050$ 
& $0.7639 \pm 0.0038$ & $0.8065 \pm 0.0057$ \\

$D_{A,1}D_{B,0}$ 
& $0.3420 \pm 0.0060$ & $0.3127 \pm 0.0090$ 
& $0.2693 \pm 0.0098$ & $0.2033 \pm 0.0075$ \\
\hline
\end{tabular}
\end{table}

\section{Phase locking system}
To stabilize the relative phase between the two spatial qubits, we use the optical phase-locked look (OPLL) \cite{yu2025chip,thomas2024path}. In our setup, one on-chip MMI (purple box in Fig. \ref{supp1}(b)) and the off-chip PD serve as the phase detector, the PLL control board provides the feedback, and the PS acts as the optical voltage-controlled oscillator. 

We observe an interference contrast above 15~dB, indicating good temporal overlap of the two optical pulses. Without phase locking, the interference signal fluctuates on a timescale of about 10~s. After phase locking is enabled, the measured power remains stable at $-51.2 \pm 0.2$~dBm, corresponding to a relative power fluctuation of $\pm 4.6\%$. This confirms that the relative phase between the two fibers is well locked. Although occasional unlock events occur, the system typically relocks within 2~s. During the experiment, data acquisition is paused during these brief events to avoid affecting the results.

\end{document}